\documentstyle[preprint,eqsecnum,aps]{revtex}
\begin{document}
\draft
\tighten
\preprint{\vbox{
\rightline{ZU-TH-98/20}   
\rightline{SU-GP-98/10-1}
\rightline{gr-qc/9812024}}}
\title{On the Generality of Refined Algebraic Quantization}
\author{Domenico Giulini}
\address{Institute for Theoretical Physics, CH-8057 Z\"urich, Switzerland}
\author{Donald Marolf}
\address{Physics Department, Syracuse University, Syracuse, New York 13244}
\date{November, 1998}
\maketitle

\begin{abstract}

The Dirac quantization `procedure' for constrained systems is well known
to have many subtleties and ambiguities.  Within this ill-defined
framework, we explore the generality of 
a particular interpretation of the Dirac procedure known as
refined algebraic quantization.  We find
technical conditions under which refined algebraic quantization 
can reproduce the general implementation of the Dirac scheme 
for systems whose constraints
form a Lie algebra with structure constants.
The main result is that, under appropriate conditions, the choice
of an inner product on the physical states is {\it equivalent} to the choice
of a ``rigging map'' in refined algebraic quantization.

\end{abstract}

\vfil
\eject

\baselineskip = 16pt
\section{Introduction}

The Dirac quantization scheme \cite{Dirac} has long been a subject of 
study by mathematical physicists, by researchers interested in quantum 
gravity, and by others.  Briefly, it suggests that certain equations 
of motion (the constraints of a gauge system) need not be imposed 
directly at the operator level, but should be imposed as conditions 
that select certain `physical states.'  This procedure, however, has 
a number of difficulties and ambiguities, not least of which arise
from the fact that additional input is typically required to define a Hilbert 
space structure on the physical states.  Some progress toward resolving 
these issues (or at least toward structuring their discussion) has been 
made in each of the BRST approach (see, for example \cite{BRS,T,HT}), the
Geometric Quantization approach \cite{woodhouse},
coherent state quantization \cite{K1,K2}, $C^*$-algebra methods \cite{G1,G2},
the Algebraic 
Quantization approach \cite{AA,AT,GMM}, the Klein-Gordon style approach (in 
the context of minisuperspace models of cosmology) \cite{KG}, and the 
Refined Algebraic Quantization approach \cite{ALMMT,BC} (which is 
closely related to Rieffel Induction \cite{KL,R} and other works 
\cite{AH,others}).

Some researchers will feel that Refined Algebraic Quantization (RAQ)
is the most abstract of these attempts.  This is to be expected of a truly
general approach, and 
we argue below that RAQ is to some degree universal. 
Since the Dirac procedure is well known to be rather vaguely defined, we 
must first spend a certain amount of time motivating and stating a precise
version of the Dirac procedure.  One of the main inputs that we will make
will be to assume that the $*$-algebra  ${\cal A}_{\rm obs}$
of observables is fixed and represented on a dense, linear subspace $\Phi$ 
of some (non-physical) Hilbert space before the constraints 
are solved.  We also argue in section IIA that it is natural to impose a 
number of technical conditions, the most important of which are that
1) the constraints are solved in the space $\Phi^*$ of linear functionals
on $\Phi$, 2) the physical   Hilbert space ${\cal H}_{\rm phys}$
is in fact contained in the space $\Phi^*$, and 3) that the Hilbert
space topology of ${\cal H}_{\rm phys}$ is finer than the topology induced
on ${\cal H}_{\rm phys}$ by inclusion in $\Phi^*$.   At the end
of section IIA, we state a precise version of the Dirac scheme\footnote{Note
that we do not claim this to be the {\it only} precise formulation of
the Dirac scheme.} within which we can investigate the generality
of refined algebraic quantization.  

This paper is concerned with the mathematical aspects of constraint
quantization.  Although it is self-contained, it is not intended to
be a review.  More introductory treatments of RAQ including a number
of worked examples are available in \cite{ALMMT,BC}.
The examples considered in \cite{others,BIX,BDT,PI,LR} may also be helpful.
In particular, they illustrate in detail the relevance of RAQ to particular
issues in gravitational physics.

Below, we will consider only systems whose constraints form a Lie algebra, 
as it is only for such systems that RAQ has yet been formulated.
We note however that our main motivation, and the original one for RAQ, comes
from gravity.  The full gravitational constraints form an algebra that involves
structure functions as opposed to structure constants, and so is not at present
addressed even in principle by RAQ.  Our hope is that RAQ can be generalized
to include cases with structure constants.  This is expected to be difficult,
due to the difficulties discussed in \cite{KK}.  However, \cite{GM} takes
a tantalizing
step in this direction in its treatment of non-unimodular Lie groups.

A skeletal discussion of refined algebraic quantization itself
is given in section IIB, at which point it is shown that all quantizations
produced by RAQ belong to the class described 
in section IIA.  This sets the stage
for the discussion of section III, in which the generality of
refined algebraic quantization is investigated within the framework of
section IIA.  It turns out that there are two main stumbling blocks
to proving that refined algebraic quantization can reproduce an
arbitrary quantization allowed by section IIA.  One
of these has to do with the domains of unbounded operators.  It can
be overcome by an extra technical assumption which, unfortunately, is
not entirely satisfactory.  The status of this assumption is discussed
in section \ref{final}.  The other stumbling block has
to do with showing that the full physical Hilbert space can be
obtained, and not just a nontrivial subrepresentation.   As discussed
in section \ref{final}, this can be
overcome by a physically motivated (but somewhat complicated) assumption.  
In the end, we do arrive at a set of conditions under which RAQ can reproduce
a general implementation of the interpretation of
the Dirac procedure that we state in section IIA.

\section{The Setting}

In part A of this section, we review the Dirac procedure and motivate
certain choices that will be made in formulating a precise version of this
procedure.
In part B, we then show that refined algebraic quantization in fact
respects these choices.  This sets the stage for the discussion of
section III, in which we explore the generality of refined algebraic
quantization within our interpretation of the Dirac scheme.

\subsection{A precise form of the Dirac procedure.}

Briefly (but not quite as briefly as above), the Dirac scheme
introduces constraint operators $C_i$ on a linear space $L$ and 
requires `physical' states $|\psi\rangle$ to be 
solutions\footnote{There have been suggestions\cite{non-U} that, when 
the constraints generate a non-unimodular gauge group, the physical 
states should not be exactly solutions to the constraints in the 
sense of (\ref{constr}).  The suggestion is that $C_i|\psi \rangle$ 
should give $|\psi\rangle$ times a certain complex number that 
depends on the group and the particular generator $C_i$. When the 
Dirac scheme is interpreted in this way, the argument below 
continues to hold with minimal modification.} of the constraints:
\begin{equation}
\label{constr}
C_i |\psi \rangle =0.
\end{equation}
Operators that commute with the constraints are called observables, 
and these map physical states to physical states.  An inner product
should be defined on the physical states to make them into a Hilbert space
in which the observables act.  Below, we will assume that the
constraints are `first-class' in the sense of Dirac \cite{Dirac} and
that they form a Lie algebra (i.e., that their commutators close with
structure constants as opposed to structure functions).

Now, there has been much discussion of observables in the context
of constraint quantization (see \cite{others,Bryce,Karel,Chris} and 
references therein). Our philosophy 
is that the $*$-algebra of quantum observables should be fixed and
represented on a Hilbert space {\it before}
the constraints are solved\footnote{Note that one cannot simply stop at this point.
The constraints contain important physical information and must be imposed, else
the full equations of motion will not be satisfied even in the classical limit.}.  
Since the $*$-relations of this algebra
must be related to the physical inner product, this will give
us some modicum of control over the rest of the construction.
We assume that the space carrying this (not yet physical) representation is a 
dense and linear subspace of the so-called auxiliary Hilbert space 
${\cal H}_{\rm aux}$.
This idea is explicitly shared by RAQ \cite{ALMMT,BC}.
The original Algebraic Quantization \cite{AA} requires  this $*$-algebra to be
fixed, but not that it be represented on an inner product space
at this stage.

Note that we have not required the observable $*$-algebra to be
explicitly constructed in any sense.  The difficulties of explicitly
constructing observables in complicated gauge theories are well known.
It is sufficient that the $*$-algebra be defined implicitly. 
We also assume that there is some Hilbert space ${\cal H}_{\rm aux}$ 
associated with $L$ such that observables are linear but not necessarily 
bounded operators on ${\cal H}_{\rm aux}$.
In practice, however, the natural order
is usually reversed: one starts with some natural Hilbert space,
like the $L^2$-space over the unreduced configuration space
(containing gauge degrees of freedom), on which the constraints 
are represented by linear operators.
Recalling that the classical constraints are real, it  appears
natural\footnote{This is certainly the case when the constraints form a Lie
algebra.  However, when the algebra involves structure functions, taking the
$C_i$ self-adjoint is in general not consistent (see \cite{KK}).  This is
the main difficulty
in generalizing RAQ to systems with structure functions.}
to identify the $C_i$ with self-adjoint operators on
${\cal H}_{\rm aux}$ (though generically they will not be
bounded). In general, it is impossible to identify $L$ with
${\cal H}_{\rm aux}$. For example, if
the spectrum of any given constraint $C_i$ is purely continuous, 
then it has no normalizable eigenvectors, and in particular none 
with eigenvalue zero. Thus, the Hilbert space itself contains no 
solutions.  What is needed is some more general mathematical 
structure, and it is natural to take a hint from spectral theory 
in which `generalized eigenvectors' (which are not normalizable) 
can be identified as certain types of distributions (dual states) 
associated with ${\cal H}_{\rm aux}$. Typical examples of this sort 
are the constructions of position- and momentum-`eigenstates' in 
ordinary quantum mechanics.  
It may of course also happen that zero is simply not part of the whole 
spectum. In this case normal spectral theory does not offer any 
solutions. But still this does not yet imply that no `solutions' to the 
constraints exist. Since they require to pick a particular representation 
of the gauge group generated by the $\{C_i\}$ from among all the 
representations `contained' in that on ${\cal H}_{\rm aux}$, one can indeed 
considerably weaken the notion of `containment' to solve the constraints 
in many such problematic cases. This is further explained in \cite{GM}. 
This reference also explains why and how the Dirac condition \ref{constr} 
has to be modified in case the group generated by the $\{C_i\}$ is not 
unimodular, in which case the spectral value zero is not relevant.

Following this suggestion we proceed by looking for a dense linear 
subspace $\Phi\subset{\cal H}_{\rm aux}$ with the properties that
(i) $\Phi$ is contained in the domain of each $C_i$, and
(ii) $\Phi$ is left invariant (as a set) under the action of each $C_i$.
In order to define the algebra of observables we now look for the set
of {\it all} operators $A$ on ${\cal H}_{\rm aux}$ such that
(i') the domain of $A$ and $A^{\dagger}$ contains $\Phi$, and 
(ii') $A$ and $A^{\dagger}$ leave $\Phi$ invariant. This set contains the
$C_i$'s and forms a $*$-algebra, ${\cal A}$, of linear operators on
$\Phi$. Within ${\cal A}$ we may consider the commutant $\{C_i\}'$,
i.e. the set of all $A\in{\cal A}$ commuting with every $C_i$, which is
a $*$-subalgebra. This is essentially our $*$-algebra of observables,
${\cal A}_{\rm obs}$. 
 Note that ${\cal A}_{\rm obs}$ becomes an algebra only {\it after}
restriction to $\Phi$. This is because we allowed for unbounded 
operators so that multiplication as operators on ${\cal H}_{\rm aux}$ 
makes generally no sense due to mismatches between ranges and domains.
The $*$-operation is that induced by the adjoint operation $\dagger$
for operators on ${\cal H}_{\rm aux}$, which on ${\cal A}_{\rm obs}$ 
becomes an abstract $*$ operation which we denote by a $\star$. 
{}From now on this operation will be the only carrier of information 
about the inner product of ${\cal H}_{\rm aux}$.
We also remark on the fact that the subalgebra generated
by the $C_i$ might have a nontrivial intersection with
${\cal A}_{\rm obs}$, given by the center $\{C_i\}^c:=\{C_i\}\cap\{C_i\}'$
of $\{C_i\}$, whose elements annihilate physical states. But physical
states are also annihilated by any product of operators in
${\cal A}_{\rm obs}$ which contain at least one member from $\{C_i\}^c$,
i.e., by the ideal $I$ in ${\cal A}_{\rm obs}$ generated by $\{C_i\}^c$.
Hence we should eventually identify the algebra of physical observables
with the quotient ${\cal A}_{\rm obs}/I$. This being understood, we shall
save notation and just consider ${\cal A}_{\rm obs}$.

The purpose of considering $\Phi$ is to introduce the concept of
distributional solutions to the constraints. We recall
(see, e.g. \cite{Bohm}) that to make the notion of distributional
states precise requires the Hilbert space structure of
${\cal H}_{\rm aux}$ {\it and} the choice of of some linear subspace
$\Phi \subset {\cal H}_{\rm aux}$.  The space $\Phi$ often comes equipped 
with its own intrinsic topology, which is to be
distinguished from the one that it inherits from being a
subspace of ${\cal H}_{\rm aux}$.  The intrinsic topology is usually 
finer and complete.\footnote{Standard examples of Gel'fand-triples in 
quantum mechanics have as $\Phi$ the space of rapidly decreasing or 
compactly supported $C^{\infty}$-functions. These spaces are topologized
by a complete, nuclear topology.} Here, however, we will simply take
$\Phi$ to have the topology induced by inclusion in ${\cal H}_{\rm aux}$.
Without loss of generality we took $\Phi$ to be dense in ${\cal H}_{\rm
aux}$, as the auxiliary Hilbert space can always be replaced by
the closure of $\Phi$.

Naturally associated with $\Phi$ is its algebraic dual, $\Phi^*$, 
given by the space of all linear maps from $\Phi$ into the complex
numbers.  Here we follow a weaker construction than that of standard
spectral analysis, as we use the algebraic dual instead of the
topological dual (which would contain only continuous maps).
The topology of $\Phi^*$ 
is that of pointwise convergence, which means that a
sequence $f_n\in\Phi^*$ converges to $f\in\Phi^*$ if and only if
$f_n(\phi)\rightarrow f(\phi)$ for all $\phi\in\Phi$. The point
now is that ${\cal H}_{\rm aux}$ can be naturally identified
with a subspace of $\Phi^*$, since vectors in ${\cal H}_{\rm aux}$ 
define linear functionals on the subset $\Phi$ by taking 
inner products. Hence there is a natural inclusion 
$j:{\cal H}_{\rm aux}\hookrightarrow\Phi^*$.   

Here again, the intrinsic
topology of ${\cal H}_{\rm aux}$ (given by the Hilbert norm) should be 
distinguished from that inherited as a subspace of $\Phi^*$. The former 
one is finer so that the inclusion $j$ is continuous. In summary, 
we have the following triple of spaces and continuous embeddings
(with a little abuse of language we may call it a `Gel'fand triple')
\begin{equation}
\label{sequence}
\Phi               \mathop{\hookrightarrow}^i
{\cal H}_{\rm aux} \mathop{\hookrightarrow}^j
\Phi^*.
\end{equation}

Depending on the choice of $\Phi$, $\Phi^*$ should be thought of as  
`a controlled enlargement' of ${\cal H}_{\rm aux}$. Technically, the 
enlargement may be understood as completion in a coarser 
topology. It is now the space of 
distributional states $\Phi^*$ that we identify with $L$
and amongst which we seek the solutions to the constraints.
If zero is not in the discrete part of the spectrum of all $C_i$, 
solutions must be sought in the complement of ${\cal H}_{\rm aux}$ in 
$\Phi^*$. We do not address here the interesting question of how the 
choice of $\Phi$ is to be made in general
but merely note that some scheme along the
lines outlined above appears necessary for 
the implementation of our philosophy. 
We strongly suspect that the choice of $\Phi$ requires some form of 
physical (and not just mathematical) input.  
See \cite{BC} for some 
discussion on this subject, with examples.

Having established the various spaces, we next need to understand the
action of operators. Since the constraints preserve $\Phi$ they have
a dual action on $\Phi^*$ (displayed in (\ref{dual})) and we may
discuss distributional solutions
$f \in \Phi^*$, defined by
\begin{equation}
\label{sol}
f(C\phi)=0 \qquad\hbox{for all}\ \phi\in\Phi.
\end{equation}   
Recall that the observable algebra should act on the space of 
solutions. 
Clearly ${\cal A}_{\rm obs}$ acts
on $L$ since an operator ${\cal O}$ on ${\cal H}_{\rm aux}$ has a
natural action on $L = \Phi^*$ if and only if its adjoint
${\cal O}^\dagger$ maps the space $\Phi$ into itself.
In this case, ${\cal O}$ can take any $f\in\Phi^*$ to ${\cal O}f$,
defined by the dual action of ${\cal O}$ on $\Phi$:
\begin{equation}
\label{dual}
({\cal O}f)(\phi) := f({\cal O}^\dagger \phi)\qquad\hbox{for all}\ 
\phi\in\Phi,
\end{equation}
where for ${\cal O}\in{\cal A}_{\rm obs}$ we should write $\star$ 
instead of $\dagger$. 
It is due to the adjoint (i.e., $\star$) in (\ref{dual}) that $\Phi^*$ 
will carry an {\it anti}-linear representation of ${\cal A}_{\rm obs}$.

The final issues have to do with the topology on $\Phi^*$ and the 
construction of the physical Hilbert space ${\cal H}_{\rm phys}$.  
The most general construction one could imagine would be to identify 
some subspace in the space ${\cal V }\subset\Phi^*$ of all solutions 
to (\ref{constr}) and make it into a pre-Hilbert space by introducing 
an inner product.  One would then complete this to a Hilbert space 
${\cal H}_{\rm phys}$.  This completion, however, could result in a 
vastly larger physical Hilbert space.  To control this, we
wish every physical state to represent 
a genuine solution in the sense of (\ref{sol}), which means that
we require ${\cal H}_{\rm phys}\subseteq {\cal V}$. 
Note that the identification of ${\cal H}_{\rm phys}$ with a subset of
${\cal V}$ is on the level of sets only. The topology on 
${\cal H}_{\rm phys}$ is that of norm convergence and not the subspace 
topology induced by $\Phi^*$. In particular, in the former topology 
${\cal H}_{\rm phys}$ must be complete. Hence it is natural to assume 
that the Hilbert space topology be finer than that induced by $\Phi^*$, 
that is, if $f_n$ converges to $f$ in ${\cal H}$, then $f_n$ converges to 
$f$ in $\Phi^*$ as well.

The observables should act via an (anti-)$*$-representation on 
${\cal H}_{\rm phys}$. 
 If ${\cal H}_{\rm phys}$ were all of ${\cal V}$, and not just a 
subspace, then each operator together with its adjoint would be defined 
everywhere on ${\cal H}_{\rm phys}$ and hence necessarily be 
bounded. In this case we would have obtained an (anti-)$*$-representation 
of ${\cal A}_{\rm obs}$ into the $*$-algebra of bounded operators. 
Presently it is not obvious to us what precisely the technical 
restrictions are that would ensure the  possibility to turn all 
of ${\cal V}$ into the physical Hilbert space, and whether 
appropriate adjustments in one's choice of $\Phi$ could lead to such 
a possibility. Hence we proceed by admitting that ${\cal A}_{\rm obs}$
is generally represented by unbounded operators, which means that there 
is also no reason for the observables to preserve ${\cal H}_{\rm phys}$.
This implies that we must take care to state in just what sense this
algebra is represented on ${\cal H}_{\rm phys}$.  We proceed in analogy with
our treatment of ${\cal A}_{\rm obs}$ on ${\cal H}_{\rm aux}$.  That is, 
we assume that there is some dense subspace $\Phi_{\rm phys} \subset
{\cal H}_{\rm phys}$ such that $\Phi_{\rm phys}$ is a common invariant
domain for all operators ${\cal O}$ in ${\cal A}_{\rm obs}$.  This allows
the algebraic relations to be represented in the action of ${\cal A}_{\rm obs}$
on $\Phi_{\rm phys}$.  We also require that the star relations be
represented in the sense that, for any $\phi_1,\phi_2 \in \Phi_{\rm phys}$ and
any ${\cal O} \in {\cal A}_{\rm obs}$, 
we have
\begin{equation}
(\phi_1, {\cal O} \phi_2) = \overline{ (\phi_2, {\cal O}^\star \phi_1)      }
\end{equation}
where $\star$ represents the $*$-operation in the algebra.  In this sense, 
the subspace $\Phi_{\rm phys} \subset {\cal H}_{\rm phys}$ carries
a star representation of the observable algebra.
By construction this automatically holds if ${\cal H}_{\rm phys}={\cal V}$, 
and is put as an extra requirement on ${\cal H}_{\rm phys}$ if it is chosen 
as a proper subspace.

Following the idea that the $*$-algebra of observables
is derived from operators of an auxiliary Hilbert space and fixed 
before the constraints are solved 
has now led us to a well-defined
setting in which to discuss the Dirac procedure.  We have a Hilbert
space ${\cal H}_{\rm aux}$, a set of self-adjoint
constraint operators $C_i$, and a subspace $\Phi$ which is mapped to 
itself by the constraints.  Without loss of generality, we may take
$\Phi$ to be dense in ${\cal H}_{\rm aux}$. 
The observable algebra ${\cal A}_{\rm obs}$ is defined by the set
of operators on ${\cal H}_{\rm aux}$ which, after restriction to $\Phi$,
commute with the constraints and map $\Phi$ to itself.
We seek a Hilbert space ${\cal H}_{\rm phys}$ which can be identified 
(as a vector space representation of ${\cal A}_{\rm obs}$) with
a subspace of $\Phi^*$, the algebraic dual to $\Phi$.  
Under this identification, 
states in ${\cal H}_{\rm phys}$ are annihilated by the (dual) action of
the constraints on $\Phi^*$, and a dense subspace $\Phi_{\rm phys}$ carries
an (anti)-$*$-representation of the 
observable algebra given by its dual action on $\Phi^*$.  Finally, 
the Hilbert space topology of ${\cal H}_{\rm phys}$ must be finer
than the topology on $\Phi^*$.

\subsection{Refined Algebraic Quantization}

Let us now take a moment to recall what is meant by refined algebraic
quantization, and to show that it fits into the precise form of the Dirac
procedure stated above.
First, we recall \cite{ALMMT,BC} that Refined Algebraic
Quantization requires a choice of ${\cal H}_{\rm aux}$, $C_i$, 
and $\Phi$ and gives a definition of ${\cal A}_{\rm obs}$ in exactly 
the same manner as section IIA. The main point of RAQ, however, is 
that the inner product is defined by a so-called `rigging map' 
$\eta$ from $\Phi$ to $\Phi^*$ through
\begin{equation}
\label{ip}
(\eta(\phi_1), \eta(\phi_2) )_{\rm phys} := \phi_1[\eta(\phi_2)] := 
\eta(\phi_2)[\phi_1], 
\end{equation}
where the physical Hilbert space, ${{\cal H}'}_{\rm phys}$, is now 
defined by taking the closure of the image of $\eta$ in this inner 
product. Alternatively, one may read (\ref{ip}) as a
semi-definite inner product on $\Phi$ and identify 
${{\cal H}'}_{\rm phys}$ with the closure of $\Phi/\hbox{kernel}(\eta)$.
The rigging map is required to satisfy certain properties (it must be
`real', `positive', and `symmetric') that are equivalent to the inner
product (\ref{ip}) being Hermitian.

The rigging map must also intertwine the
representations of the observable algebra on $\Phi$ and  $\Phi^*$.
In \cite{ALMMT,BC}, this last property was 
called `commuting' with the observable algebra.
This means that, for any ${\cal O} \in {\cal A}_{\rm phys}$
and any $\phi \in \Phi$, we have
\begin{equation}
{\cal O}( \eta \phi) = \eta ( {\cal O} \phi).
\end{equation}
Note that this implies that the image of $\eta$ (all of which lies
in ${\cal H}'_{\rm phys}$) is an invariant domain for the action of
${\cal A}_{\rm obs}$ on ${\cal H}'_{\rm phys}$.  Thus, we may
take the image of $\eta$ to be the space $\Phi_{\rm phys}$ required
by the formulation of the Dirac procedure stated at the end of
section IIA.

We would like to show that refined algebraic quantization in fact
satisfies all requirements of our version of the Dirac procedure.
At this point, it is clear that the requirements
for ${\cal H}_{\rm aux}$, $C_i$, $\Phi$, and ${\cal A}_{\rm obs}$ are
satisfied, but we must still check that, in RAQ,  
${\cal H}'_{\rm phys}$ may be considered a subspace of $\Phi^*$ with a
finer topology.
To verify this, consider the map $\sigma: {\cal H}'_{\rm phys}
\rightarrow \Phi^*$ defined for $f\in {\cal H}'_{\rm phys}$ by
\begin{equation}
\label{sigma}
(\sigma f)[\phi] = (f, \eta(\phi))_{\rm phys}.
\end{equation}
We first note that $\sigma f$ vanishes only if $f$ is orthogonal to
all states in the image of $\eta$.  But this image is dense in 
${\cal H}'_{\rm phys}$ by construction, 
so $\sigma$ is an embedding of linear spaces.  Moreover, 
$\sigma$ is continuous since the inner product (\ref{sigma}) is
continuous in $f$.  Thus, we may use $\sigma$ to identify 
${\cal H}'_{\rm phys}$ with a subspace of $\Phi^*$ and we see that the Hilbert
space topology of ${\cal H}'_{\rm phys}$ is finer than the topology induced
by this inclusion.  Moreover, since $\eta$ commutes with the action of
${\cal A}_{\rm obs}$, we have
\begin{eqnarray}
(\sigma {\cal O} f)[\phi] & = & (f, \eta({\cal O}^\dagger \phi))_{\rm phys}
\cr
&=& (\sigma f) [{\cal O}^\dagger \phi ] \cr
&=& ({\cal O} \sigma f) [\phi ]. 
\end{eqnarray}
 Thus, $\sigma:\hbox{Image}(\eta)\rightarrow\hbox{Image}(\sigma)$ is an 
isomorphism of vector space representations of ${\cal A}_{\rm obs}$. 
If follows that refined algebraic quantization does in fact satisfy all of 
the requirements stated at the end of section IIA.

\section{The Generality of Refined Algebraic Quantization}

The last paragraph in section IIA gives a precise statement
of the Dirac prescription but, a priori, there may still be a great
many distinct ways to construct appropriate physical Hilbert spaces, which
may or may not be related to RAQ.
However, we now show that our version of the Dirac procedure (from IIA)
provides at least a natural candidate
for a rigging map.  Whether or not it actually is a rigging map
depends on certain technical conditions involving the domains of operators.
In the case that this condition is satisfied, it will follow that
refined algebraic quantization can construct at least a nontrivial
subrepresentation of the original representation of ${\cal A}_{\rm obs}$
on ${\cal H}_{\rm phys}$.  Under a further technical condition, the
entire representation of ${\cal A}_{\rm obs}$ on ${\cal H}_{\rm phys}$
can be constructed through RAQ.

As observed in section IIB, 
refined algebraic
quantization and section IIA both involve a choice of 
${\cal H}_{\rm aux}$, $C_i$, 
and $\Phi$, and the restrictions on this choice are the same in both
cases.  They then use identical
definitions of ${\cal A}_{\rm obs}$.   However, section IIA 
places only a few restrictions on the physical Hilbert space ${\cal H}_{\rm
phys}$, while refined algebraic quantization requires that the physical
Hilbert space be constructed by a certain method involving a rigging
map.  Thus,
our task is to investigate whether a general physical inner product
satisfying the conditions of section IIA can in fact be constructed
through a rigging map as would be required by refined algebraic quantization.

To do this, consider any $\phi \in \Phi$. 
Since ${\cal H}_{\rm phys} \subset \Phi^*$ and since convergence in 
${\cal H}_{\rm phys}$
implies convergence in $\Phi^*$, $\phi$ defines a continuous linear
functional on ${\cal H}_{\rm phys}$.  But ${\cal H}_{\rm phys}$ 
is a Hilbert space, which
means that there is some $\phi_0 \in {\cal H}_{\rm phys}$ such that, 
for all $f \in {\cal H}_{\rm phys}$, $\phi(f) = (\phi_0, f)$.  Thus, we may
introduce a map $\eta : \Phi \rightarrow {\cal H}_{\rm phys}$ given by
$\eta(\phi) = \phi_0$.

The map $\eta$ is {\it anti-}linear, and the inner product on 
${\cal H}_{\rm phys}$, restricted to $\hbox{Image}(\eta)$, 
is 
\begin{equation}
\label{ip2}
(\eta(\phi_1), \eta(\phi_2) )_{\rm phys} = \phi_1[\eta(\phi_2)] = 
\eta(\phi_2)[\phi_1].
\end{equation}
If $\eta$ is a rigging map, this is just
the Refined Algebraic inner product.  Note that since the inner product
on ${\cal H}_{\rm phys}$ is Hermitian, $\eta$ is real, positive, and 
symmetric (in the sense of Refined Algebraic Quantization).  

Now, it remains only to investigate the behavior of $\eta$ with 
respect to the observables.  Here, a difficulty arises involving the domains
of the unbounded operators in ${\cal A}_{\rm obs}$.  Let $\Phi'{}_{\rm phys}$
be the image of $\eta$ in ${\cal H}_{\rm phys}$.  Refined algebraic
quantization requires that $\Phi'{}_{\rm phys}$ carry a $*$-representation of
${\cal A}_{\rm obs}$.  However, section IIA in no way restricts the 
choice of which subspace should carry such a $*$-representation, and in 
particular does not appear to restrict the domains of the operators in
any way.  Thus, it does not appear necessary that $\Phi'{}_{\rm phys}$
even overlap the domain of any observable ${\cal O}$.  Note that this
would not be an issue if we could somehow arrange to work with 
bounded operators on the physical Hilbert space,  as in the case where 
${\cal H}_{\rm phys}={\cal V}$.

Thus, in order to proceed, we must introduce the additional assumption
that the image $\Phi'{}_{\rm phys}$ of $\eta$ is contained in
the dense subspace $\Phi_{\rm phys}$.  In this case, we can can show
that $\eta$ satisfies the last requirement of a rigging map.  Namely, it
commutes with the action of observables.  To see this, consider any
$\phi_1, \phi_2 \in \Phi$ and ${\cal O} \in {\cal A}_{\rm phys}$. 
We can use (\ref{ip2}), the hermiticity of the physical inner product,
and the definition of the dual action to conclude
that
\begin{equation}
 \eta ({\cal O} \phi_2) [\phi_1] = \overline{\eta(\phi_1) [
{\cal O} \phi_2] } = \overline {{\cal O}^\star\eta(\phi_1) [\phi_2]}.
\end{equation}
But this is the (complex conjugate of) the inner product
$(\eta \phi_2, {\cal O}^\star \eta(\phi_1))_{\rm phys}$ on the physical
Hilbert space.  Now, using the fact that $\Phi_{\rm phys}$ carries
a $*$ representation of the observables, we can rewrite the right-hand
side above as $(\eta \phi_1, {\cal O}\eta (\phi_2))_{\rm phys}
= ({\cal O} \eta(\phi_2)) [\phi_1]$.  Thus, the observables do indeed
commute with the map $\eta$ and it is a rigging map.
The closure of the image of $\eta$ in ${\cal H}_{\rm phys}$
is the corresponding Hilbert space ${\cal H}'{}_{\rm phys}$
defined by RAQ.  Note that since ${\cal A}_{\rm obs}$ maps $\Phi$ into itself, 
it must also preserve the image of $\eta$.  

Now consider any nonzero $f \in {\cal H}_{\rm phys}$. Then, clearly,
there exists some $\phi \in \Phi$ for which $f(\phi)$ and hence 
$\eta(\phi)$ is non-zero. Thus, the image of $\eta$ is not trivial 
unless ${\cal H}_{\rm phys} = \{ 0 \}$.

It follows that, given our extra condition about the containment of
$\Phi'{}_{\rm phys}$ in $\Phi_{\rm phys}$, 
at least a nontrivial subrepresentation of
${\cal A}_{\rm obs}$ on ${\cal H}_{\rm phys}$
can be constructed
by Refined Algebraic Quantization for
any implementation of the Dirac
scheme satisfying the requirements of section IIA.
If $\Phi'{}_{\rm phys}$ is dense in ${\cal H}_{\rm phys}$, then RAQ
can construct the entire representation.  Requiring this space
to be dense in ${\cal H}_{\rm phys}$ may be natural from a physical
point of view as it guarantees that, in some sense,
any state in ${\cal H}_{\rm phys}$
can be approximated arbitrarily well by a state in the auxiliary
space\footnote{In fact, it seems natural to require that $\Phi$
approximate ${\cal H}_{\rm phys}$ in the sense of states on (perhaps
$C*$-subalgebras of) ${\cal A}_{\rm obs}$.  The consequences of such
a requirement will be left for future studies.}.

\section{Discussion}
\label{final}

For systems whose constraints form a Lie algebra, 
we have shown that RAQ is, to some degree, universal.  However, we
have introduced a number of technical conditions whose physical meaning is
unclear, and we have arrived only at the conclusion that RAQ is capable
of constructing a nontrivial subrepresentation of the physical algebra.

One might expect that one could dispense with some of the technical conditions
by working with bounded operators instead of unbounded ones.  While this
is true to some extent, our current understanding is that the main effect
would simply be to replace the technical conditions above with technical
conditions of other sorts.  The main point here is that we have
found no natural way to ensure that some topology (or norm)
placed on the observable
algebra at the level of ${\cal H}_{\rm aux}$ survives in any way
to the level of ${\cal H}_{\rm phys}$.  Thus, restricting the observable
algebra to bounded operators on ${\cal H}_{\rm aux}$ does not appear
to guarantee (in, for example, refined algebraic quantization)
that this algebra will be represented by bounded operators
on ${\cal H}_{\rm phys}$.  In particular, the powerful theorems concerning
$C*$-algebras do not apply here, as the restriction that ${\cal A}_{\rm obs}$
leave $\Phi$ invariant leads us to expect that ${\cal A}_{\rm obs}$ is
not complete.  Requiring by fiat that the entire algebra ${\cal A}_{\rm obs}$
be represented by bounded operators on ${\cal H}_{\rm phys}$ is
a strong restriction, and may be too strong to be physically acceptable.
An example will be presented in \cite{GM} in the context of non-unimodular
groups where it is argued that the physically correct approach maps
certain operators on ${\cal H}_{\rm aux}$ to operators with a {\it larger}
norm on ${\cal H}_{\rm phys}$.  In particular, it will map a unitary
representation $U(g)$ of the gauge group to the (non-unitary)
representation $\Delta^{1/2}(g)$ given by the modular homomorphism
$\Delta: G \rightarrow {\bf R}^+$.  Thus, for the moment at least, 
we feel constrained to work with unbounded operators.

Now, as stated above, even under all of these technical conditions we
have only been able to show that RAQ can reproduce a nontrivial 
subrepresentation of ${\cal A}_{\rm obs}$ on ${\cal H}_{\rm phys}$.
However, we will now show that the introduction of a further assumption
guarantees that RAQ can reproduce the entire representation of the
observables.  While this new assumption is a bit awkward to state, we
will see that it follows from a physical motivation.

Suppose that ${\cal H}_{\rm phys} = {\cal H}_1 \oplus {\cal H}_2$.  Recall
that if ${\cal H}_1$ separately reduces [i.e., is invariant under]
${\cal A}_{\rm obs}$, then so does ${\cal H}_2$. Hence any matrix 
element of an operator in ${\cal A}_{\rm obs}$ between a state 
from ${\cal H}_1$ and a state from ${\cal H}_2$ necessarily vanishes,
so that there is no interference between states in ${\cal H}_1$ and
${\cal H}_2$ with respect to ${\cal A}_{\rm obs}$. 
One says \cite{ss} that there is a superselection rule between 
the two sectors ${\cal H}_1$ and ${\cal H}_2$. 
Pure physical states must therefore lie either in ${\cal H}_1$ or 
${\cal H}_2$, which form physical Hilbert spaces in their own 
right.\footnote{A superposition of a vector in ${\cal H}_1$ 
with a vector in ${\cal H}_2$ still defines a physical state, in the 
sense that it defines a positive linear functional on ${\cal A}_{\rm obs}$.
But because of the absence of interference terms, this functional is 
the same as the mixture of the corresponding individual density 
matrices. Hence it is a mixed physical state.} 

Now, it may happen that such a superselection rule arises in a physical
Hilbert space ${\cal H}_{\rm phys}$ constructed along the lines
of section II.  Having constructed the direct sum ${\cal H}_1
\oplus {\cal H}_2$ in this way, it is tempting to try to construct
each sector separately following the Dirac approach.  However,  it
is not clear that this is in general possible.  Suppose then that
the sector ${\cal H}_2$ cannot be constructed in isolation while
${\cal H}_1$ can. If
one is to insist upon a Dirac approach, then it is natural to
consider ${\cal H}_2$ less physical than ${\cal H}_1$.

We may therefore wish to add the extra physical assumption that
every superselected sector can be constructed {\it separately}
by a Dirac procedure as defined in section II.  Note
that this condition is satisfied by RAQ, as one may construct
${\cal H}_2$ simply by replacing $\Phi$ by 
the subspace $\Phi_2$ that is mapped into ${\cal H}_2$ by
the rigging map.   This new assumption
will in fact be sufficient to show that the entire physical
Hilbert space ${\cal H}_{\rm phys}$ can be constructed
from RAQ.  To see this, simply consider the direct sum ${\cal H}_\Sigma$
of all sectors in ${\cal H}_{\rm phys}$ which can be constructed
through RAQ.  ${\cal H}_\Sigma$ carries a 
representation of ${\cal A}_{\rm obs}$, 
so there is a superselection rule between ${\cal H}_\Sigma$ and its orthogonal
complement.  Now, by our new assumption, this orthogonal complement
can be constructed separately following a Dirac procedure.  But if this
complement is nontrivial, then
by our earlier result it must contain some nontrivial subrepresentation 
constructible by RAQ.  Since this is false, we must have
${\cal H}_\Sigma = {\cal H}_{\rm phys}$. 
  
Thus, the choices available in the Dirac scheme are perhaps not
as large as might be thought.  Although the final results are not
as conclusive as one might like, we have managed to shed light on
the class of implementations of the Dirac procedure which can be
obtained by refined algebraic quantization.  This will be useful
in interpreting the strength of
results that can be derived within the mathematically
powerful setting of refined algebraic quantization.  
In particular, this will provide a useful perspective
on the result to be derived in
\cite{GM}.  That work also considers systems whose constraints form
a Lie algebra and
shows that, when they converge properly, group averaging 
techniques \cite{ALMMT,AH} give the unique implementation of the 
Dirac quantization scheme in the above sense. 

\acknowledgments

The authors wish to thank Jim Hartle for prompting them to consider
the generality of Refined Algebraic Quantization.  DM was supported
in part by NSF grant PHY97-22362 and by funds provided by Syracuse 
University. DG by the Albert-Einstein-Institute, the Swiss National 
Science Foundation and the Tomalla Foundation.

\end{document}